\documentclass[%
superscriptaddress,
showkeys,
preprint,
 amsmath,amssymb,
 aps,
prc
]{revtex4-1}

\usepackage{graphicx}
\usepackage{dcolumn}
\usepackage{bm}
\usepackage{booktabs}
\usepackage{soul, color, xcolor}
\usepackage{hyperref}

\usepackage{CJKutf8}
\begin{document}
\preprint{APS/123-QED}

\title{Effects of incompressibility on the neutron-proton equilibration in $^{70}$Zn + $^{70}$Zn collisions at 35 MeV/nucleon}

\author{Erxi Xiao}
\affiliation{Sino-French Institute of Nuclear Engineering and Technology, Sun Yat-sen University, Zhuhai 519082, China}
\author{Yu Yang}
\affiliation{Sino-French Institute of Nuclear Engineering and Technology, Sun Yat-sen University, Zhuhai 519082, China}
\author{Yingge Huang}
\affiliation{Sino-French Institute of Nuclear Engineering and Technology, Sun Yat-sen University, Zhuhai 519082, China}
\author{Zhen Zhang}
\affiliation{Sino-French Institute of Nuclear Engineering and Technology, Sun Yat-sen University, Zhuhai 519082, China}
\author{Long Zhu}
\affiliation{Sino-French Institute of Nuclear Engineering and Technology, Sun Yat-sen University, Zhuhai 519082, China}
\author{Jun Su}\email{sujun3@mail.sysu.edu.cn} %
\affiliation{Sino-French Institute of Nuclear Engineering and Technology, Sun Yat-sen University, Zhuhai 519082, China}
\affiliation{China Nuclear Data Center, China Institute of Atomic Energy, Beijing 102413, China}
\affiliation{Guangxi Key Laboratory of Nuclear Physics and Nuclear Technology, Guangxi Normal University, Guilin 541001, China}


\date{\today}

\begin{abstract}
\begin{description}
\item[Background]
The primary goal of studying isospin dynamics via heavy-ion reactions is to explore the isospin dependence of effective interactions within the nuclear equation of state (EOS).
\item[Purpose]
This work aims to investigate the effects of nuclear incompressibility (\( K_0 \)) on neutron-proton equilibration in projectile-like fragments (PLFs).
\item[Method]
We simulate \(^{70}\)Zn + \(^{70}\)Zn collisions at 35 MeV/nucleon using the isospin-dependent quantum molecular dynamics (IQMD) model, coupled with the statistical decay code GEMINI.
\item[Results]
The IQMD simulations not only reproduce experimental data patterns but also reveal the dynamic mechanisms underlying the binary breakup of PLFs. 
The rotation of PLFs is influenced by the transformation of angular momentum, which is connected to the isoscalar component of the EOS. 
This connection explains why shifts in \( K_0 \) affect the description of neutron-proton equilibration as measured by PLF rotation. 
The simulations demonstrate that a model with a smaller \( K_0 \) paired with a softer symmetry energy, or a larger \( K_0 \) with a slightly stiffer symmetry energy, both offer better indications of neutron-proton equilibration.

\item[Conclusion]
Considering the uncertainty in \( K_0 \), the slope of the symmetry energy is constrained within the range of \( L = 20 \sim 40 \) MeV, providing valuable insights into the nuclear equation of state.
\end{description}
\end{abstract}

\maketitle
\section{\label{int}Introduction}
The nuclear equation of state (EoS), as a representation of the nuclear force, has a wide-ranging impact on both terrestrial and astrophysical phenomena, including isotopic abundances, cluster formation at low densities, the outer crust of neutron stars, and the formation of heavy elements in supernova explosions \cite{lattimer2001neutron, janka2007theory,moller2012new}. Two key parameters in the EoS are the incompressibility \( K_0 \) and the symmetry energy, both of which are crucial for understanding the behavior of nuclear matter.

The incompressibility \( K_0 \), which quantifies the response of symmetric nuclear matter to compression at saturation density, is a fundamental parameter in the isoscalar part of the EoS. Several experimental methods have been used to constrain \( K_0 \), including measurements of giant monopole resonances \cite{garg2018compression, shlomo2006deducing, su2020constraints}, fusion reaction cross sections \cite{zanganeh2012dynamical, rana2024nuclear}, nucleon elliptic flows \cite{danielewicz2002determination, wu2023probing, cozma2018feasibility}, kaon production \cite{fuchs2001probing}, and observations of neutron stars and their mergers \cite{lattimer2007neutron, perego2022probing}. Theoretical studies, both relativistic and non-relativistic, have converged on a consensus value of \( K_0 = 230 \pm 40 \, \text{MeV} \) \cite{garg2018compression, colo2014symmetry, li2021differential, khan2012constraining}.

The symmetry energy, which reflects the energy difference between pure neutron matter and symmetric nuclear matter, plays a crucial role in the isovector part of the EoS. Significant progress has been made in constraining its density dependence through measurements of neutron skin thickness \cite{brown2000neutron,centelles2009nuclear,zhang2013constraining}, isospin diffusion and drift \cite{baran2005reaction,li2008recent,colonna2020collision,su2020isospin}, charged pion ratios \cite{tsang2017pion,xu2024comparing}, and the mass-radius relations of neutron stars \cite{abbott2017gw170817,fattoyev2018neutron,zhang2018combined,xie2019bayesian}. A comprehensive review synthesized results from 28 different models, combining data from terrestrial nuclear experiments and astrophysical observations. It provided values of \( S_0 = 31.6 \pm 2.7 \, \text{MeV} \) for the symmetry energy at saturation density and \( L = 58.9 \pm 16 \, \text{MeV} \) for the slope of the symmetry energy \cite{li2019towards}.

Neutron-proton equilibration has been extensively studied in heavy-ion collisions at Fermi energies \cite{tsang2009constraints,piantelli2020dynamical,camaiani2021isospin,ciampi2023quasiprojectile,fable2024isospin,hudan2012tracking,stiefel2014symmetry,jedele2017characterizing,rodriguez2017detailed,jedele2023investigating}. The time scale of this equilibration provides a sensitive probe for constraining the density dependence of the symmetry energy \cite{stiefel2014symmetry,hudan2012tracking}. In particular, studies have measured a neutron-proton equilibration time scale of around \( \sim 0.3 \, \text{zs} \) \cite{jedele2017characterizing,rodriguez2017detailed}. Both the Constrained Molecular Dynamics (CoMD) and Antisymmetrized Molecular Dynamics (AMD) models indicate that a softer symmetry energy provides a better description of this equilibration time scale, with constraints on the slope parameter \( L \) falling below 51 MeV and 21 MeV, respectively \cite{jedele2023investigating}.

Several theoretical studies have demonstrated that the uncertainty in the incompressibility \( K_0 \) should be considered when constraining the density dependence of the symmetry energy \cite{long2024effects,zhang2015covariance,zhang2020constraints,xu2021bayesian,kumar2021incompressibility}. Calculations using the Boltzmann-Uehling-Uhlenbeck model have shown that charged pion production is sensitive to both \( K_0 \) and the slope of the symmetry energy \cite{long2024effects}. This sensitivity arises because the incompressibility influences the maximum attainable density in heavy-ion collisions, which in turn affects pion production \cite{reisdorf2007systematics}. As a result, a correlation between the incompressibility and symmetry energy parameters emerges when charged pion production is used to extract information about high-density symmetry energy \cite{xu2024comparing}.

A linear relationship between \( K_0 \) and the slope of the symmetry energy \( L \) was found through Bayesian uncertainty quantification, based on data from giant monopole resonances \cite{xu2021bayesian}. This relationship can be understood by considering how the centroid energy of nuclei relates to density fluctuations near the saturation point \cite{colo2004microscopic,colo2014symmetry}. Therefore, the uncertainty in \( K_0 \) must be accounted for to extract reliable information about the symmetry energy from neutron-proton equilibration experiments.

This work focuses on the effects of incompressibility \( K_0 \) on neutron-proton equilibration, as measured by the rotation of projectile-like fragments (PLFs), and constrains the slope of the symmetry energy while accounting for the uncertainty in \( K_0 \).
The paper is organized as follows: Section \ref{model} describes the theoretical framework. Section \ref{results} presents the results and discussion, while Section \ref{summary} provides the conclusions.

\section{\label{model}Theoretical framework}
The wave function for each nucleon in the IQMD model is represented by a Gaussian wave packet
     \begin{equation}	
          \phi _\mathit{i} (\mathbf{r},\mathit{t}  )= \frac{1}{(2\pi \mathit{L})^{3/4}} e^{-\frac{[\mathbf{r}-\mathbf{r_\mathit{i}(\mathit{t} ) }  ]^2}{4L} }e^{\frac{\mathit{i}\mathbf{r}\cdot \mathbf{p}_\mathit{i}(\mathit{t} )    }{\hbar } },
     \end{equation}
where $\mathbf{r}_{i}$ and $\mathbf{p}_{i}$ represent respectively the average position and momentum of the $i$-th nucleon, and $L$ is related to the extension of the wave packet.
The total $N$-body wave function is assumed to be the direct product of these coherent states.
Through a Wigner transformation of the wave function, the density distribution function of a system is given by
\begin{equation}
\rho(\boldsymbol{r})=\frac1{(2\pi L)^{3/2}}\sum_{i=1}^N\mathrm{e}^{-[\boldsymbol{r}-\boldsymbol{r_i}(t)]^2/2L}
\end{equation}

The propagation of nucleons in the system under the self-consistently generated mean field is governed by Hamiltonian equations of motion,
\begin{equation}
 \dot{\mathbf{r} }_\mathit{i} = \nabla _{\mathbf{p}_i }\mathit{H},  \dot{\mathbf{p}}_i = -\nabla _{\mathbf{r}_i}\mathit{H}.
\end{equation}
The Hamiltonian $H$ consists of the kinetic energy, the Coulomb potential energy and the nuclear potential energy.
The nuclear potential energy is obtained from the integration of the Skyrme energy density functional,
\begin{equation}
\begin{aligned}
V(\rho,\delta) = & \frac{\alpha}{2} \frac{\rho^2}{\rho_0} + \frac{\beta}{\gamma+1} \frac{\rho^{\gamma+1}}{\rho_0^{\gamma}} +\frac{g_{sur}}{2\rho_0 } (\nabla\rho )^2 \\
&+ \frac{C_{s}}{2}(\frac{\rho}{\rho_{0}})^{\gamma_{i}} \rho \delta ^{2},
    \label{V}
  \end{aligned}
  \end{equation}
where $\rho_0$ is the normal density and $\delta$ is the isospin asymmetry.
The corresponding incompressibility is calculated by
\begin{equation}
K_0 = -\frac{6}{5}\frac{\hbar^2}{2m}(\frac{3\pi^2}{2}\rho_0)^{2/3}+9\frac{\gamma(\gamma-1)}{\gamma+1}\beta.
    \label{K0}
\end{equation}

It should be noticed that the momentum-dependent interaction (MDI) is not included in Eq. \ref{V}.
The MDI is optional in the IQMD model, see our previous works \cite{su2016correlation,su2017effects}.
However, when the MDI is considered in model, the excited PLFs tend to decay by nucleon evaporation but not binary breakup.
The main reason is that the numerical approximation to solve the MDI lead to the spurious emission of nucleons.
How to include the MDI in the IQMD model without numerical approximation is still an open question, and the important topic in the Transport Model Evaluation Project \cite{xu2016understanding,xu2024comparing,zhang2018comparison,ono2019comparison,colonna2021comparison}.

The form of the symmetry potential energy density, shown as the fourth term in Eq. \ref{V}, was proposed by Tsang $et$ $al$. \cite{tsang2009constraints}.
The symmetry energy is composed of the kinetic energy from the Fermi motions of nucleons and the symmetry potential as
\begin{equation}
E_{sym} (\rho ) = \frac{1}{3} \frac{\hbar ^2}{2m} (\frac{2}{3}\pi ^2\rho  )^{2/3}+\frac{C_{s} }{2}(\frac{\rho}{\rho_0})^{\gamma _i} .
    \label{Esym}
\end{equation}
We can obtain the symmetry energy at saturation density $E_{sym} (\rho_0 )=13+\frac{C_{s} }{2}$ from Eq. \ref{Esym}.
The slope of the symmetry energy at the saturation density is calculated by
\begin{equation}
L = 26+\frac{3}{2} C_s\gamma _i.
    \label{L}
\end{equation}
The $E_{sym} (\rho_0 )=32$ MeV is considered.
The parameters used in present work are listed in table \ref{table}.
\setlength{\tabcolsep}{1.0em}
\begin{table}[]
\renewcommand{\arraystretch}{0.8}
\caption {\label{table}
The parameters of the nuclear potential energy adopted in the present work.}
\begin{tabular}
{lcccccccc}
\hline\hline
& $\alpha$  &  $\beta$  &  $\gamma$  & $g_{sur}$  & $C_s$ & $\gamma _i$ & $K_0$ & L \\
& (MeV) & (MeV) & & (MeV fm$^2$) & (MeV) & & (MeV) & (MeV) \\
\hline
S1 & -356.00  & 303.00  &  7/6  &  20.80  & 38.06 & -0.10 & 200 & 20 \\
S2 & -356.00  & 303.00  &  7/6  &  20.80  & 38.06 & 0.25 & 200 & 40 \\
S3 & -356.00  & 303.00  &  7/6  &  20.80  & 38.06 & 0.75 & 200 & 70 \\
S4 & -356.00  & 303.00  &  7/6  &  20.80  & 38.06 & 1.5 & 200 & 110 \\
H1 & -168.45  & 115.86  &  1.5  &  20.80  & 38.06 & 0.25 & 270 & 40\\
H2 & -168.45  & 115.86  &  1.5  &  20.80  & 38.06 & 0.75 & 270 & 70\\
H3 & -168.45  & 115.86  &  1.5  &  20.80  & 38.06 & 1.5 & 270 & 110\\
 \hline\hline
\end{tabular}
\end{table}

To compensate for the fermionic feature, the phase-space density constraint (PSDC) method was firstly applied in the CoMD model \cite{papa2001constrained}.
This method has also been introduced in IQMD model.
In our previous work, the PSDC method has been shown to be important in describing the central collisions at Fermi energies \cite{su2018fusion, su2023temperature} and projectile fragmentation \cite{su2018dynamical,su2019uniform}.
Based on the PSDC method, the probability of phase-space occupation $\overline{\mathit{f}_i}$ is calculated by performing the integration on a hypercube of volume $\mathit{h}^3$ in the phase space centered around the $\mathit{i}$th nucleon.
\begin{equation}
\bar{f}_i =0.621+\sum_{j\ne 1}^{N} \frac{\delta _{\tau _j,\tau _i}}{2} \int_{h^3}\frac{1}{\pi ^3\hbar ^3}e^{-\frac{(r_j-r_i)^2}{2L}- \frac{(p_j-p_i)^2}{\hbar ^2/2L}}d^3rd^3p.
\label{ith_possibility}
\end{equation}
Here 0.621 is the contribution itself and $\tau _i$ means the isospin degree of freedom.
The phase-space occupation $\overline{\mathit{f}_i}$ for each nucleon is checked at each time step.
If the value of phase-space occupation $\overline{\mathit{f}_i}$ is greater than 1, the momentum of the $\mathit{i}$th nucleon is changed randomly by the many-body elastic scattering.
Only if $\overline{\mathit{f}_i}$ and $\overline{\mathit{f}_j}$ at the final states are both less than 1, the result of many-body elastic scattering is accepted.

In order to avoid the spurious emissions of nucleons, the evolution by the IQMD model will be stopped when the excitation energies of the two heaviest prefragments are less than $E_{stop}$ = 3 MeV/nucleon.
The Ref. \cite{su2018dynamical} stated that there are spurious emissions of nucleons in the IQMD model due to the numerical fluctuations.
It means that a few nucleons will be evaporated even if the fragment is in its ground state.
As the time proceeds and for lighter nuclei, this effect becomes stronger.
If one can stop the dynamical simulation (by the IQMD model) and switch to the statistical simulation (by the GEMINI model) as soon as the process of PLFs breakup is over, then the spurious emissions will be controlled.
We have proven that the $E_{stop}$ settings is useful to reduce the spurious emissions of nucleons and thus improve the multiplicity of the intermediate mass fragments \cite{su2018dynamical, su2019uniform, su2022fluctuations, xiao2022dissipation, xiao2024uncertainties}.

The IQMD simulations will switch off when meeting $E_{stop}$ condition.
Then the deexcitation of prefragments are described by the statistical code GEMINI including the Hauser-Feshbach-type evaporation.
There are two assumptions in GEMINI code to describe secondary decay of prefragments.
One is that the density and structure of prefragments is close to the normal nuclei ones.
The other is that the properties of such fragments, in particular, the symmetry energy, level densities, and others, are correspond to the properties of normal nuclei.
The details of GEMINI are given in Ref. \cite{charity1988systematics, su2018influence}.

For the IQMD simulations, $10^{7} $ events are simulated for each parameter set in table \ref{table}.
Moreover, the impact parameters are randomly chosen from 0 to $b_{max}$ fm.
The maximum impact parameter is defined by $b_{max}$ = $1.2(A_{p}^{1/3}+ A_{t}^{1/3})$.
$A_{p}^{1/3}$ and $A_{t}^{1/3}$ are the mass numbers of the projectile and target, respectively.

\section{\label{results}Results and discussion}
\subsection{\label{reproduce}The dynamic properties of binary breakup of projectile-like fragments}

\begin{figure}
\centering
\includegraphics[width=1.0\textwidth,angle=0]{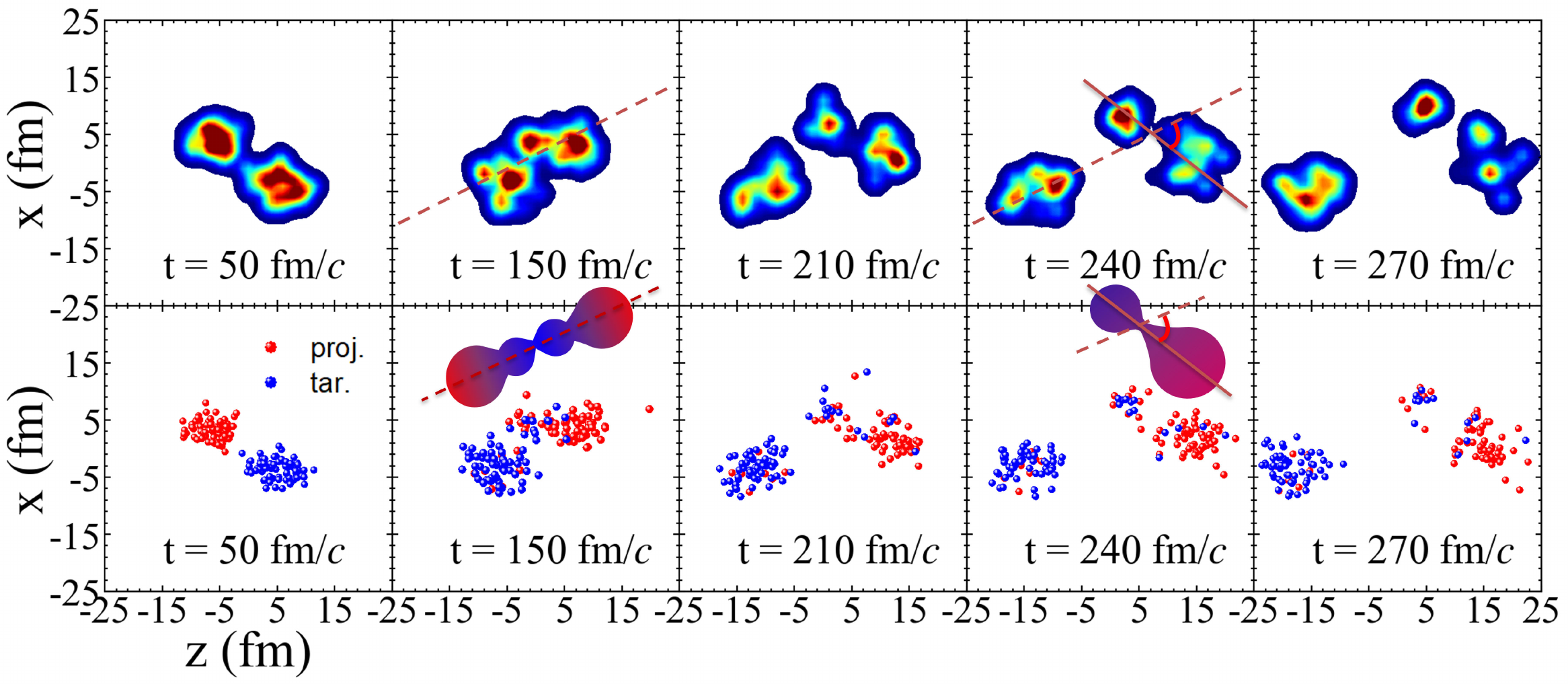}\caption{\label{1} The distributions of nucleon density (upper part) and configurational maps (bottom part), at different time steps, of one event related to the $^{70}$Zn + $^{70}$Zn collisions at 35 MeV/nucleon by IQMD model.
The separation axis between PLF and TLF is shown as dash line.
The deformation axis between HF and LF is shown as solid line.
Illustrations of dynamical deformation and isospin transport are shown in the insets.
The color denotes the composition with blue indicating neutron richness and red indicating relative neutron deficiency.
}
\end{figure}

In this subsection, we use the parameter set S2 from Table \ref{table} to focus on describing the dynamic properties of binary breakup. 
Figure \ref{1} shows the nucleon density distributions (upper panel) and configurational maps (lower panel) as a function of time for \(^{70}\)Zn + \(^{70}\)Zn collisions at 35 MeV/nucleon, simulated by the IQMD model. 
These distributions are taken from a typical event.

At 50 fm/c, the projectile and target make contact. 
By 150 fm/c, projectile-like fragment (PLF) and target-like fragment (TLF) are formed. 
Between 50 and 150 fm/c, as the developing PLF and TLF begin to separate, a low-density ``neck" forms between them. 
Neutrons flow from the developing PLF into this neck due to the density gradient, a process known as isospin drift, as the system seeks to minimize its symmetry energy \cite{colonna2020collision}.

After 150 fm/c, the PLF begins to rotate (clockwise in the figure). 
At 240 fm/c, the PLF breaks into a heavier fragment (HF) and a lighter fragment (LF) along its deformation axis (shown by the solid line). 
Since the nuclear matter from the neck is initially neutron-rich, there is an isospin asymmetry gradient within the PLF. 
From 150 to 240 fm/c, nucleons are likely to flow from the neutron-rich region (LF) to the neutron-deficient region (HF). 
This isospin diffusion drives the isospin asymmetry between the developing HF and LF towards equilibration.

The rotation of the deformed PLF is quantified by the alignment angle between the separation axis and the deformation axis \cite{jedele2017characterizing, rodriguez2017detailed, jedele2023investigating}. The separation axis is represented by the center-of-mass velocity of the PLF, \( \vec{v }_{cm} = (m_H \vec{v}_H + m_L \vec{v}_L) / (m_H + m_L) \), while the deformation axis is represented by the relative velocity between HF and LF, \( \vec{v}_{rel} = \vec{v}_H - \vec{v}_L \). 
The alignment angle is defined as:
\begin{equation}
\alpha =\arccos \frac{\vec{v }_{cm}\cdot  \vec{v }_{rel} }{\left |\vec{v }_{cm}  \right | \left |\vec{v }_{rel}  \right |} .
\label{rot}
\end{equation}

This alignment angle serves as a measure of the contact time between HF and LF \cite{hudan2012tracking, stiefel2014symmetry, jedele2017characterizing, rodriguez2017detailed, jedele2023investigating}. 
In experiments, fragments are detected with a wide angular coverage ranging from 3.6$^{\circ } $  to 167.0$^{\circ } $. 
A total charge requirement of 21 \( \leq Z_{tot} \leq \) 32 is also applied to exclude incomplete fusion events. 
Additionally, the HF and LF are required to have atomic numbers of at least \( Z_{HF} \geq 12 \) and \( Z_{LF} \geq 3 \), respectively. 
This same sorting criterion is used to select fragments in our simulation.

\begin{figure}
\centering
\includegraphics[width=0.45\textwidth,angle=0]{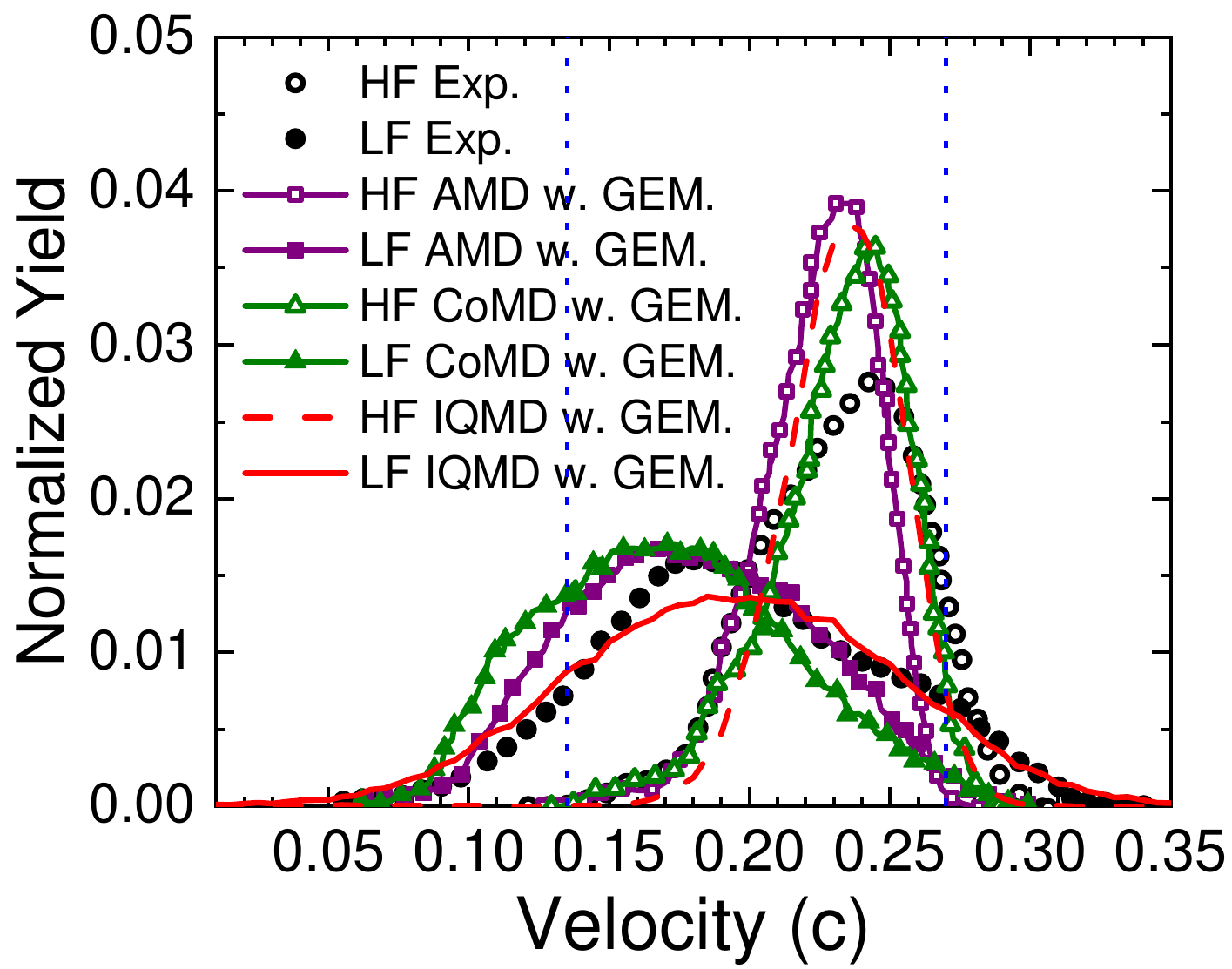}
\caption{\label{2}The longitudinal velocity distributions of HF and LF in $^{70}$Zn + $^{70}$Zn collisions at 35 MeV/nucleon.
The doted vertical lines (from left to right) correspond to the midvelocity (i.e., 0.135 $c$) and beam velocity (i.e., 0.27 $c$), respectively.
The data and calculations of AMD as well as CoMD with GEMINI models are taken in Ref. \cite{jedele2023investigating}.
}
\end{figure}

Figure \ref{2} shows the longitudinal velocity distributions of the HF and LF in \(^{70}\)Zn + \(^{70}\)Zn collisions at 35 MeV/nucleon. 
The peak positions of both the HF and LF velocity distributions are higher than the mid-velocity (0.135 \(c\)) but lower than the beam velocity (0.27 \(c\)). 
This indicates that both the HF and LF result from the binary breakup of the PLF.

The peak velocity of the HF is higher than that of the LF, suggesting that they originate from different regions. 
The LF comes from the neck region, while the HF is formed farther away from the neck, within the PLF itself. 
Due to the strong damping of nuclear matter in the neck region, the nucleons in this area experience greater dissipation of incident energy, which is converted into excitation energy. 
This leads to stronger dynamic fluctuations in the LF, explaining why the LF has a broader velocity distribution compared to the HF.

Simulations from three different transport models all produce similar velocity distributions to the experimental data. 
However, compared to the data, the calculated velocity distributions from these models are generally shifted to the left, with the exception of the LF distribution calculated by the IQMD model. 
This shift suggests that the interaction between nucleons is overestimated in these transport models.

\begin{figure}
\centering
\includegraphics[width=0.45\textwidth,angle=0]{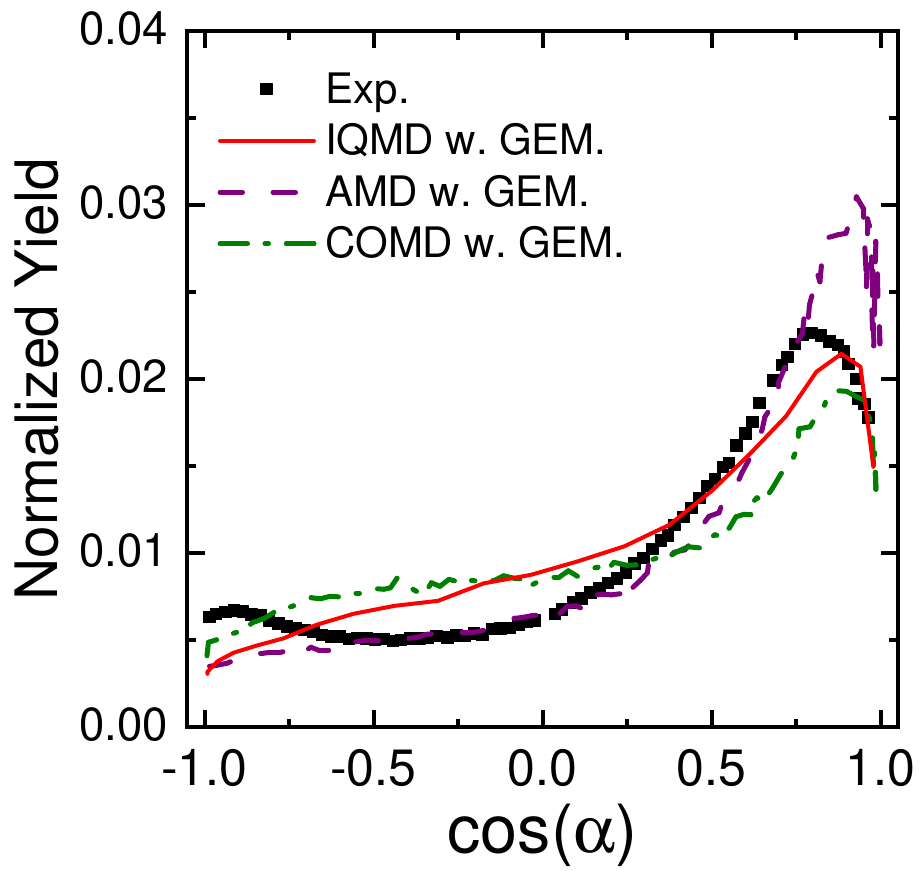}
\caption{\label{3} Experimental and simulated distributions of the alignment angle $\alpha$ in $^{70}$Zn + $^{70}$Zn collisions at 35 MeV/nucleon.
The data is shown as closed square.
The solid line, dash line, and dash-dot line represent the  calculations of IQMD, AMD, and CoMD with GEMINI models, respectively.
The data and calculations of AMD as well as CoMD with GEMINI models are taken in Ref. \cite{jedele2023investigating}.
}
\end{figure}

Figure \ref{3} shows the distributions of the alignment angle \( \alpha \) in \(^{70}\)Zn + \(^{70}\)Zn collisions at 35 MeV/nucleon. 
The yield is high and follows a Gaussian distribution in the region \(0.5 < \cos(\alpha) < 1\), indicating that the PLF tends to break apart on a short timescale, with a preference for the HF being emitted forward relative to the LF. 
In contrast, the yield is lower and shows a flattened distribution in the region \(-1 < \cos(\alpha) < 0\), suggesting that the PLF survives longer and reaches equilibrium, with HF and LF separating at random angles. 
The coexistence of short- and long-timescale breakups results in the overall distribution of \( \cos(\alpha) \) seen in Fig. \ref{3}.

All three transport models capture the general trend of the experimental data, showing a decreasing yield as the alignment angle increases. 
However, there is a noticeable underprediction of the yield at \( \cos(\alpha) = -1 \) in the simulations by all three models compared to the data, suggesting that the simulated total angular momentum is lower than observed experimentally.
The IQMD model predicts a higher yield for short-timescale breakups and a lower yield for long-timescale breakups compared to the CoMD model, but lower and higher, respectively, than the AMD model. 
These differences among the three models may be attributed to the stopping conditions applied during the dynamic stage of the simulations. 
The AMD and CoMD simulations are stopped at 300 fm/c and 1000 fm/c, respectively \cite{jedele2023investigating}, while the evolution time in IQMD simulations depends on the excitation energies of the two heaviest prefragments and typically ranges from 300 to 500 fm/c.

\begin{figure}
\centering
\includegraphics[width=0.45\textwidth,angle=0]{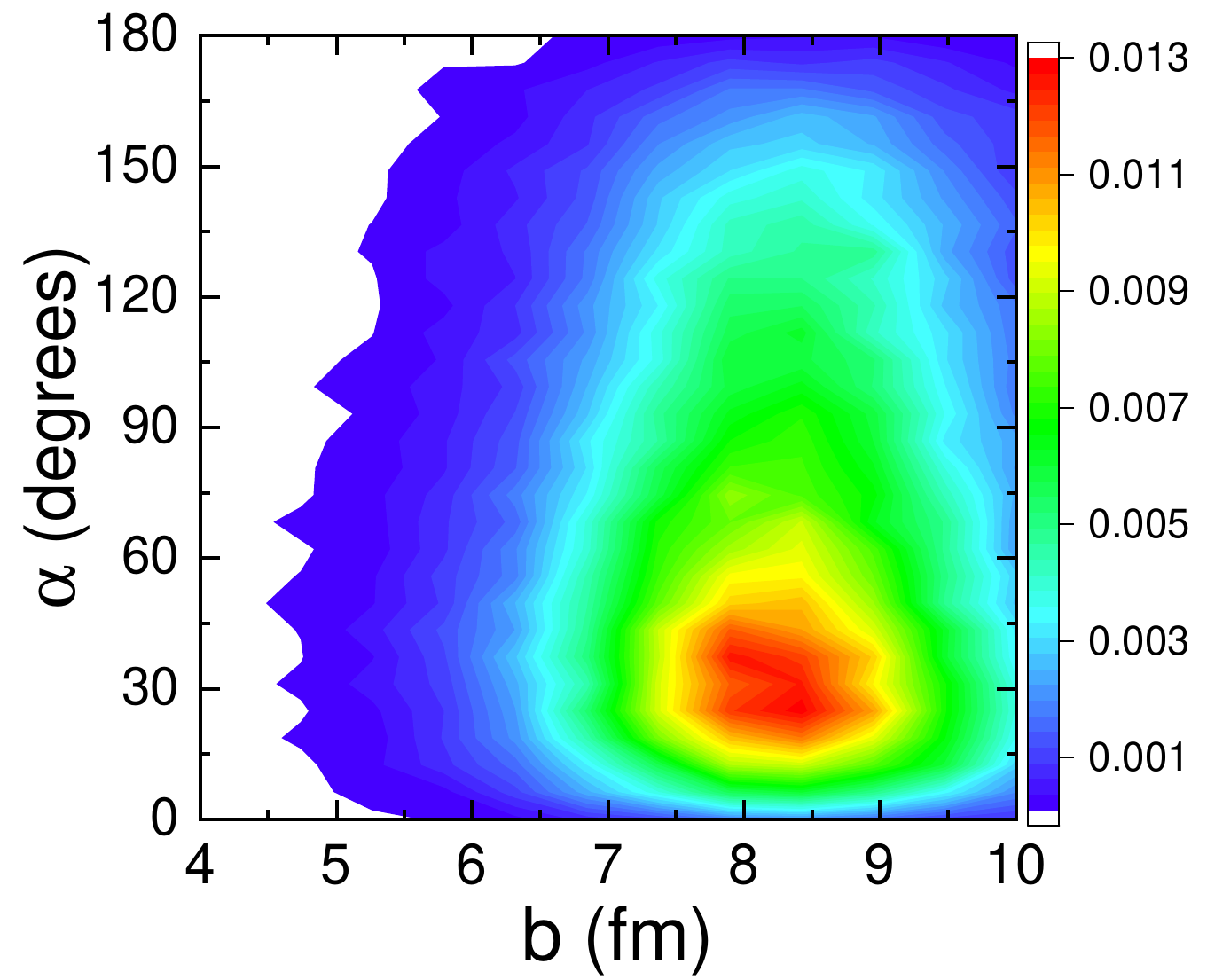}
\caption{\label{4}The alignment angles versus the impact parameters in $^{70}$Zn + $^{70}$Zn collisions at 35 MeV/nucleon by the IQMD model.}
\end{figure}

Figure \ref{4} shows the alignment angles as a function of impact parameters in \(^{70}\)Zn + \(^{70}\)Zn collisions at 35 MeV/nucleon, calculated using the IQMD model. 
The results indicate that the breakup of the PLF into the HF and LF predominantly occurs in peripheral collisions, around \( b \approx 8.5 \) fm. 
It is well known that the dynamic properties of PLFs are strongly influenced by the impact parameter, which is expected to determine the relative velocity between the HF and LF.
However, the data show no clear correlation between the alignment angle and the impact parameter. 
For a given impact parameter, the yield initially increases with the alignment angle in the range \( \alpha < 60^{\circ} \), then decreases when \( \alpha > 60^{\circ} \). 
This distribution is similar to that observed in Figure \ref{3}, supporting the reliability of the alignment angle as a measure of the contact time between the HF and LF.

\begin{figure}
\centering
\includegraphics[width=0.45\textwidth,angle=0]{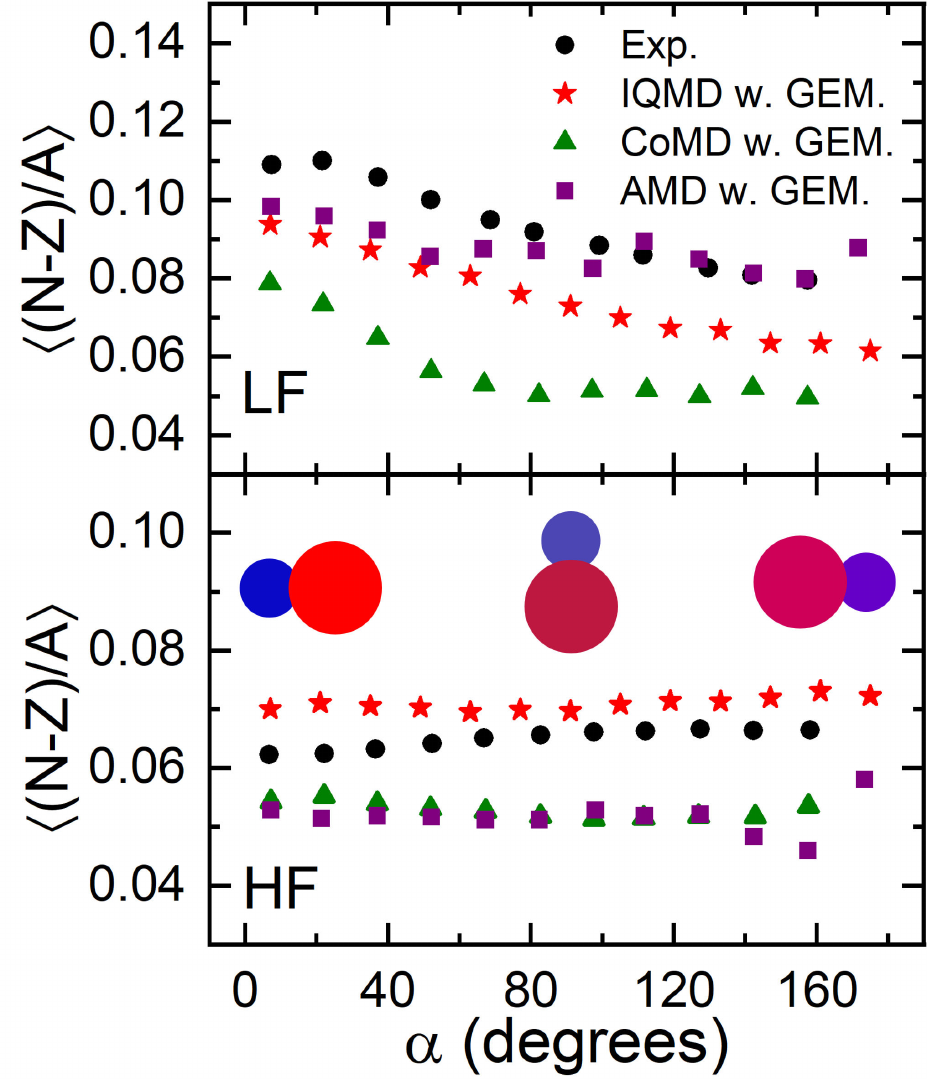}\caption{\label{5} The mean isospin asymmetry of the LF (upper panel) and HF (lower part) as function of the alignment angle in $^{70}$Zn + $^{70}$Zn collisions at 35 MeV/nucleon.
The data and calculations by the CoMD and AMD models are taken in Ref. \cite{jedele2023investigating}.
The illustration of isospin diffusion is shown in the insets.
The color denotes the composition with blue indicating neutron richness and red indicating relative neutron deficiency.
}
\end{figure}

The positive correlation between the alignment angle and contact time provides an opportunity to study time-dependent neutron-proton equilibration. Figure \ref{5} shows the mean isospin asymmetry of the LF and HF as a function of the alignment angle.
For the LF, a large initial isospin asymmetry is observed at \( \alpha = 0^{\circ} \) (t = 0 fm/c). 
As the alignment angle increases, the isospin asymmetry of the LF decreases. 
In contrast, the HF starts with a small isospin asymmetry, which increases and eventually levels off as the alignment angle grows. 
This initial isospin asymmetry is attributed to isospin drift, where neutrons accumulate in the low-density neck region, from which the LF originates. 
As the alignment angle increases, corresponding to a longer contact time between the LF and HF, the isospin asymmetry between the two fragments approaches equilibrium due to isospin diffusion. 
Isospin diffusion causes neutrons to flow from the LF to the HF, as illustrated in the insets of Fig. \ref{5}. 

The AMD and CoMD models best reproduce the data when the slope of the symmetry energy is set to \( L = 21 \) MeV (AMD) and \( L = 51 \) MeV (CoMD), respectively \cite{jedele2023investigating}. 
These calculations are shown in Fig. \ref{5}. 
For the LF, the curvature of the isospin asymmetry predicted by the AMD model is smaller than the data, while the CoMD model predicts a larger curvature. 
The curvature of the IQMD model, with \( L = 40 \) MeV, agrees with the experimental data. 
This difference between the simulations likely arises from the different choices for the slope of the symmetry energy.

Additionally, the isospin asymmetries of both the LF and HF are underestimated by the AMD and CoMD models, suggesting that these models overestimate the emission of free neutrons or neutron-rich light fragments during the preequilibrium stage. 
In contrast, the IQMD model underpredicts the isospin asymmetry of the LF but overpredicts that of the HF. 
This discrepancy may be linked to the influence of momentum-dependent interactions. 
Notably, the AMD and CoMD models include momentum-dependent interactions, while the IQMD model does not, which may account for the differences observed in comparison to the data.

\subsection{\label{drift}The effect of isospin drift on the isospin asymmetry of PLFs}

\begin{figure}
\centering
\includegraphics[width=0.45\textwidth,angle=0]{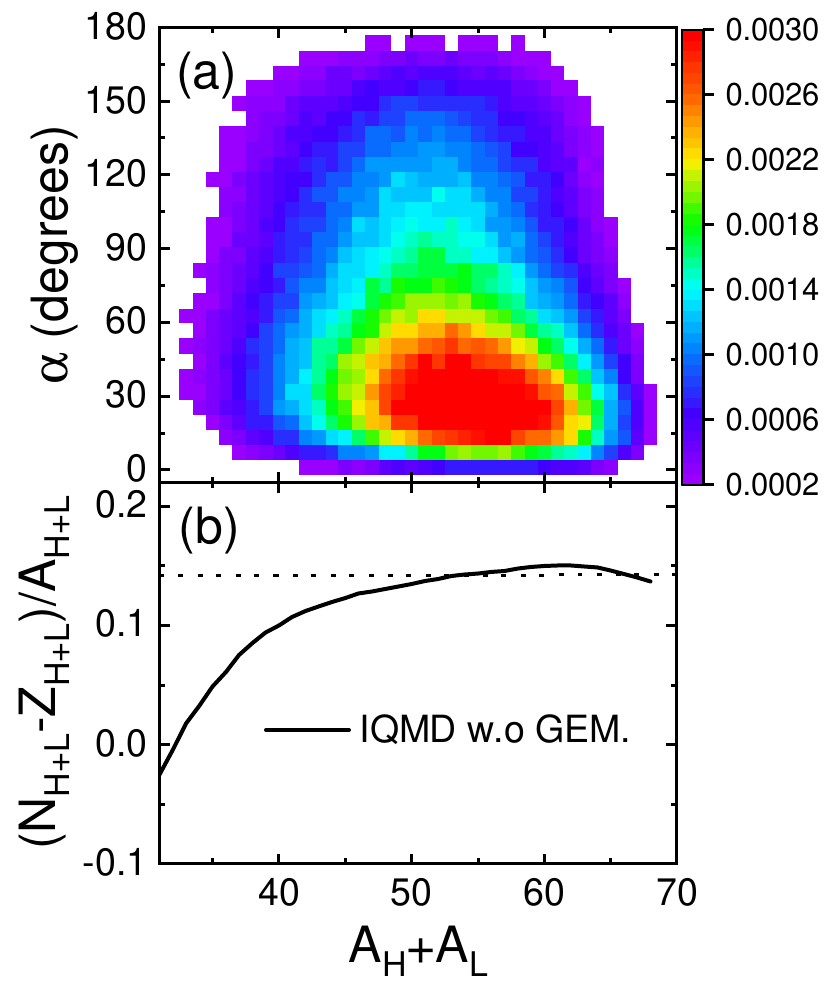}\caption{\label{6}(a) The alignment angles $\alpha$, and (b) isospin asymmetry of the HF $+$ LF system as function of total mass number both HF and LF $A_{H} + A_{L}$ in $^{70}$Zn + $^{70}$Zn collisions at 35 MeV/nucleon by IQMD model.}
\end{figure}

In heavy-ion collisions at Fermi energies, isospin observables are influenced by both isospin diffusion and drift. 
Isospin drift operates over a longer timescale and can persist throughout the entire process, from early dynamic emission to late-stage statistical evaporation \cite{zhang2017long,su2020isospin}. 
While Fig. \ref{5} demonstrates the occurrence of isospin diffusion between the HF and LF, it remains unclear whether isospin diffusion is the dominant factor driving changes in isospin asymmetry between these fragments. 
The emission of free neutrons or neutron-rich light fragments due to isospin drift could also contribute to changes in isospin asymmetry. 
This contribution can be assessed by examining the isospin asymmetry of the combined HF + LF system, calculated as \( \left \langle \frac{N_{H+L} - Z_{H+L}}{A_{H+L}} \right \rangle \).

Figure \ref{6} presents the alignment angle \( \alpha \) and isospin asymmetry \( \left \langle \frac{N_{H+L} - Z_{H+L}}{A_{H+L}} \right \rangle \) as functions of the total mass number \( A_H + A_L \). 
Since the HF and LF are primarily produced during the dynamical decay of the PLF, most events occur in the region of \( \alpha < 90^{\circ} \). 
When sorted by \( A_H + A_L \), the events are concentrated in the region where \( A_H + A_L > 45 \).

It is noteworthy that \( \left \langle \frac{N_{H+L} - Z_{H+L}}{A_{H+L}} \right \rangle \) increases rapidly with increasing \( A_H + A_L \), eventually approaching the isospin asymmetry of the projectile \( ^{70} \text{Zn} \) (0.142) for \( A_H + A_L > 50 \). This suggests that the emission of free neutrons or neutron-rich light fragments due to isospin drift decreases as \( A_H + A_L \) increases and becomes negligible when \( A_H + A_L > 50 \).

\begin{figure}
\centering
\includegraphics[width=0.45\textwidth,angle=0]{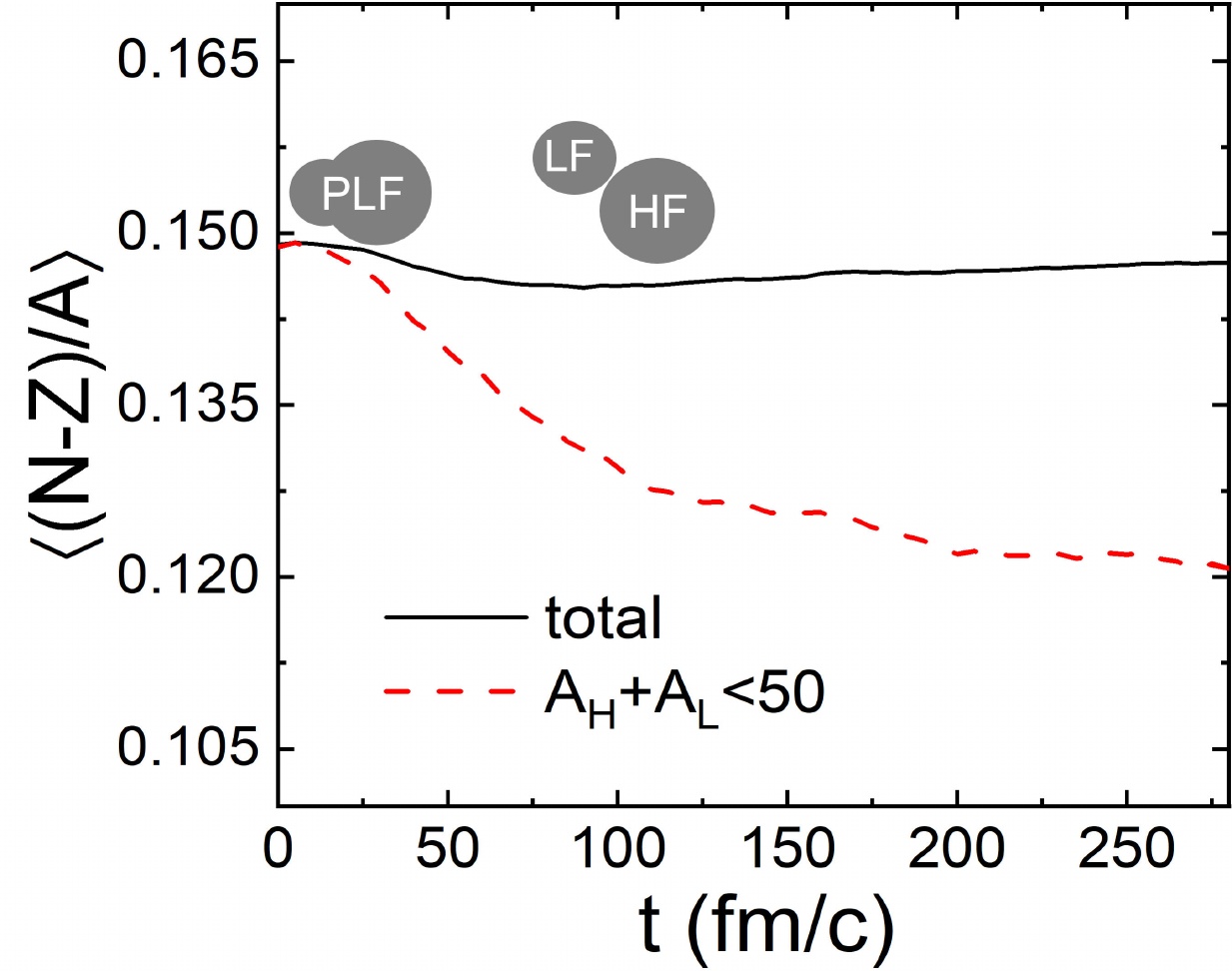}\caption{\label{7} The time evolution of the isospin asymmetry of the HF $+$ LF system by the IQMD model.}
\end{figure}

We track the evolution of isospin asymmetry throughout the entire process of binary breakup in the HF + LF system, as shown in Fig. \ref{7}. 
The time \( t = 0 \) fm/c is defined as the moment when the PLFs can first be clearly identified. Typically, the PLFs have a short survival time and have broken into HF and LF by \( t = 100 \) fm/c. The events are sorted by the total mass number of the HF and LF at the final moment, \( A_H + A_L \), as in Fig. \ref{6}.

For events where \( A_H + A_L < 50 \), the influence of isospin drift is significant. 
As time progresses, the isospin asymmetry of the HF + LF system decreases sharply due to the emission of free neutrons or neutron-rich light fragments. 
However, considering the full set of events, the overall change in isospin asymmetry throughout the binary breakup process is relatively small. 
This indicates that the isospin asymmetry between the HF and LF is primarily determined by isospin diffusion rather than drift.

\subsection{\label{K}The influence of the incompressibility on the neutron-proton equilibration}

The nuclear density of the system during nuclear reactions is influenced by the incompressibility, \( K_0 \). 
In previous work \cite{su2018fusion}, it was found that a smaller \( K_0 \) value leads to greater density fluctuations in the projectile-like system during central collisions at Fermi energies. 
It is expected that the value of \( K_0 \) will also affect the dynamical properties of the neck region, as well as the deformation and rotation of the PLF in semi-peripheral or peripheral collisions.
In peripheral collisions, the system naturally generates relative angular momentum between the projectile and target. 
Our calculations show that the conversion of this relative angular momentum into the angular momentum of the PLF itself depends on the incompressibility used in the model. 
In the subsequent breakup of the PLF, the alignment angle \( \alpha \) at a given rotation time is positively correlated with the angular momentum of the PLF. 
As a result, when extracting the relationship between the alignment angle \( \alpha \) and rotation time (which corresponds to the isospin diffusion time), the effect of the incompressibility used in the model must be taken into account.

\begin{figure}
\centering
\includegraphics[width=0.45\textwidth,angle=0]{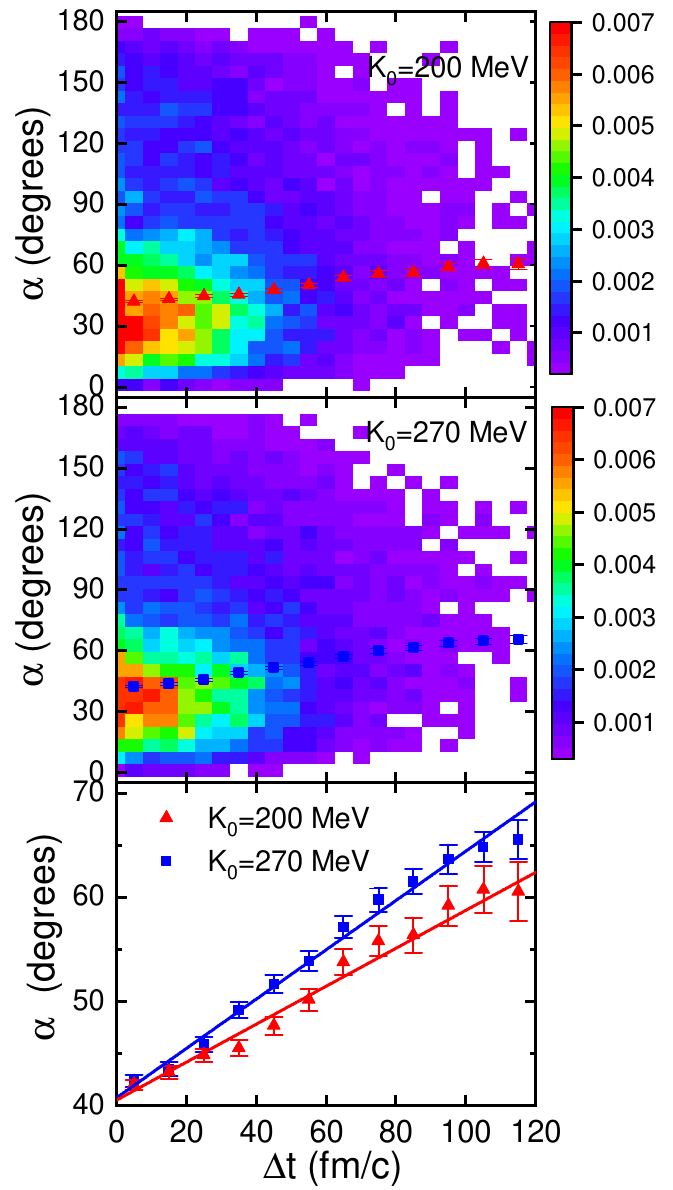}\caption{\label{8}
The alignment angle versus the rotation time of the PLF.
The cases within parameter sets S2 ($K_0$ = 200 MeV) and H1 ($K_0$ = 270 MeV) are compared.
The mean alignment angle $\left \langle \alpha  \right \rangle $ as function of the rotation time for $K_0$ = 200 and 270 MeV are plotted as triangles and squares, respectively.}
\end{figure}

The rotation time is defined as the time interval from the separation of the PLF and target-like fragment to the breakup of the PLF into the HF and LF, i.e., \( \Delta t = t_{\rm{break}} - t_{\rm{separate}} \). 
In Ref. \cite{harvey2020correlation}, CoMD simulations show that the PLF may experience a false breakup in certain events. 
This phenomenon is also observed in IQMD simulations, though it occurs less frequently.
Figure \ref{8} shows the correlation between the alignment angle and the rotation time of the PLF. The distribution of \( \alpha \) versus \( \Delta t \), shown in Figs. \ref{8}(a) and \ref{8}(b), is qualitatively consistent with results from the CoMD and AMD models \cite{harvey2020correlation, piantelli2020dynamical}.

In Figure \ref{8}(c), the average alignment angles, \( \left \langle \alpha  \right \rangle \), are plotted as a function of \( \Delta t \) for two cases: \( K_0 = 200 \) MeV and \( K_0 = 270 \) MeV. The calculations indicate that the incompressibility used in the model has a significant effect on the relationship between the alignment angle and the rotation time. 
For the same \( \Delta t \), \( \left \langle \alpha  \right \rangle \) is larger in the case of \( K_0 = 270 \) MeV compared to \( K_0 = 200 \) MeV. 
Additionally, the difference in \( \left \langle \alpha  \right \rangle \) between these two cases increases with \( \Delta t \).

\begin{figure}
\centering
\includegraphics[width=0.7\textwidth,angle=0]{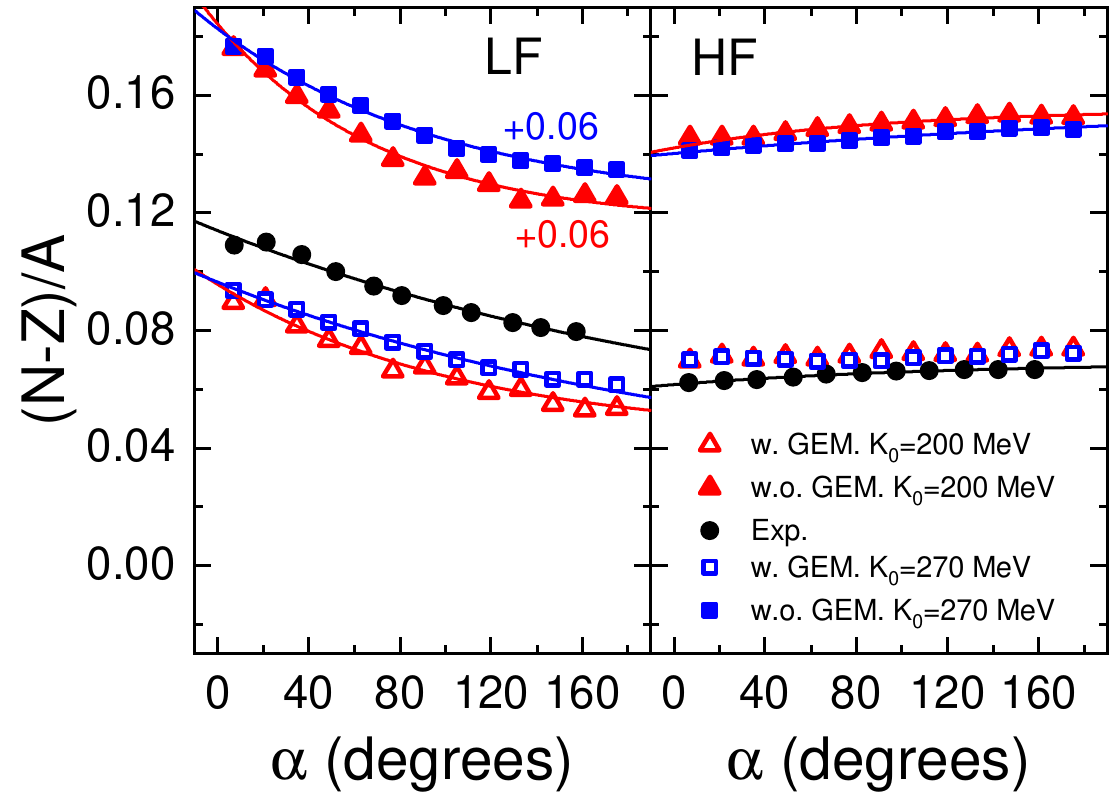}\caption{\label{9}
The mean isospin asymmetry of the LF (left panel) and HF (right part) as function of the alignment angles in $^{70}$Zn + $^{70}$Zn collisions at 35 MeV/nucleon.
The data taken in \cite{jedele2023investigating} are shown as the closed circles.
The closed and open scatters represent the IQMD and IQMD + GEMINI calculations, respectively.
The calculations with the parameter sets S2 ($K_0$ = 200 MeV) and H1 ($K_0$ = 270 MeV) are plotted as triangles and squares, respectively.}
\end{figure}

Figure \ref{9} illustrates the mean isospin asymmetry of the LF and HF as a function of the alignment angles. 
The calculations for \( K_0 = 200 \) MeV and \( K_0 = 270 \) MeV are compared to assess the influence of the incompressibility parameter used in the model on neutron-proton equilibration between the HF and LF. 
The relationship of \( \left \langle (N-Z)/A \right \rangle \) for both HF and LF versus \( \alpha \) has been fitted using an exponential form \cite{jedele2017characterizing, rodriguez2017detailed, jedele2023investigating}:
\begin{equation}
\left \langle (N-Z)/A \right \rangle = a + b \exp (-c\alpha).
\label{fit1}
\end{equation}
In this equation, \( c \) serves as a surrogate for the equilibration rate constant, providing a direct measure of the curvature of the \( \left \langle (N-Z)/A \right \rangle \) versus \( \alpha \).

The sensitivity of the equilibration rate constant \( c \) to the values of \( K_0 \) is notable. 
For the LF calculations using the IQMD model, the value of \( c \) for \( K_0 = 270 \) MeV (0.009) is smaller than that for \( K_0 = 200 \) MeV (0.014). 
A similar trend is observed for the HF, where \( c \) for \( K_0 = 270 \) MeV (0.005) is also smaller than that for \( K_0 = 200 \) MeV (0.011). 
This behavior can be explained by noting that a larger \( K_0 \) results in a steeper slope of \( \left \langle \alpha \right \rangle \) versus \( \Delta t \), which consequently leads to a flatter curvature in the \( \left \langle (N-Z)/A \right \rangle \) versus \( \alpha \) plot.

To investigate the effects of secondary decay, we compare the calculations between the IQMD model and the IQMD+GEMINI model. 
The secondary decay results in a smaller curvature for \( \left \langle (N-Z)/A \right \rangle \) of the LF versus \( \alpha \). 
For the HF, obtaining an exponential fit to the calculations by the IQMD+GEMINI model is challenging, a result that has also been noted in AMD and CoMD simulations \cite{jedele2023investigating}. Importantly, the secondary decay does not obscure the influence of \( K_0 \) on neutron-proton equilibration.
The value of \( c \) for \( K_0 = 270 \) MeV (0.005) remains smaller than that for \( K_0 = 200 \) MeV (0.009). 
Notably, the \( c \) value for \( K_0 = 270 \) MeV aligns well with the experimental \( c \) values reported in the literature \cite{jedele2023investigating}.

\begin{figure}
\centering
\includegraphics[width=0.45\textwidth,angle=0]{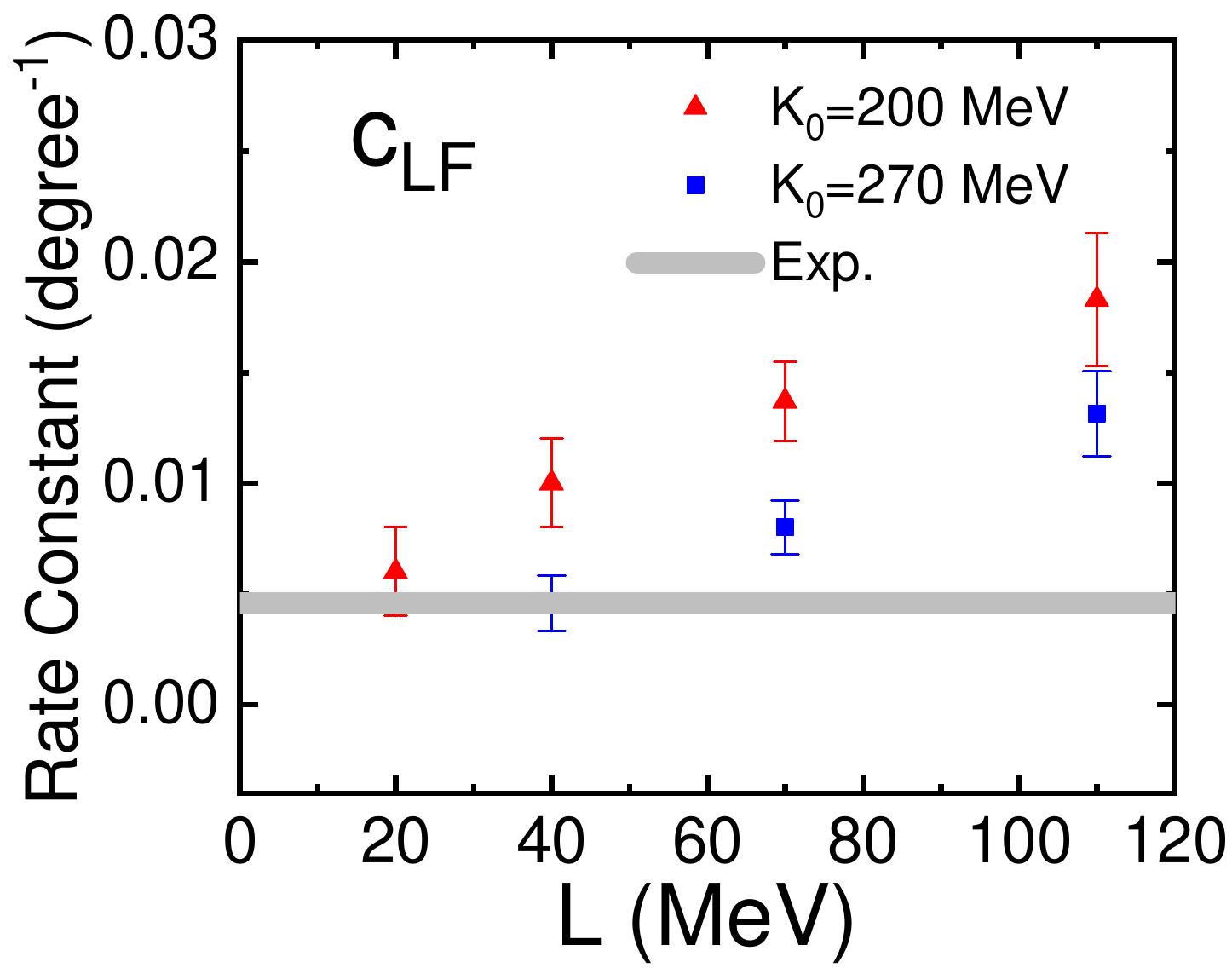}\caption{\label{10}
The $c$ for the LF as function of the slope of symmetry energy $L$ by the IQMD + GEMINI model using different parameter sets in table \ref{table}.
The calculations for $K_0$ = 200 and 270 MeV are displayed as triangles and squares respectively.
The grey marking represent the data.}
\end{figure}

The rate constant \( c \) is extracted using different parameter sets from Table \ref{table} to investigate the density dependence of symmetry energy while accounting for uncertainty in \( K_0 \). The \( c \) values for the LF obtained through the IQMD+GEMINI model are quantitatively compared to experimental \( c \), as shown in Fig. \ref{10}. 
The calculations indicate that \( c \) is sensitive to the slope of the symmetry energy \( L \), which aligns with the findings in Ref. \cite{jedele2023investigating}. 
Specifically, for both \( K_0 = 200 \) MeV and \( K_0 = 270 \) MeV, \( c \) increases with increasing \( L \). 
For \( K_0 = 200 \) MeV, the \( c \) value for \( L = 20 \) MeV is consistent with the experimental data within the error bars.

In Ref. \cite{jedele2023investigating}, the AMD and CoMD models do not consider the influence of different \( K_0 \) values on \( c \), and their calculations only provide constrained upper limits for \( L \) (i.e., \( L < 21 \) and \( L < 51 \) MeV). 
In contrast, the IQMD model demonstrates that \( c \) is also sensitive to the \( K_0 \) values: larger \( K_0 \) values yield smaller \( c \) at a given \( L \). 
Considering the uncertainty of \( K_0 \), the slope of symmetry energy is constrained to the range \( L = 20 \sim 40 \) MeV, with \( E_{\text{sym}} = 32 \) MeV fixed at the saturation point.

To intuitively understand the influence of \( K_0 \) when using the rate constant \( c \) to constrain the slope of symmetry energy, we examine the neutron-proton equilibration. 
Time-dependent Hartree-Fock calculations have been fitted using an exponential form \cite{umar2017transport, simenel2020timescales}:
\begin{equation}
 (N-Z)/A  = a+b\exp (-\frac{t}{\tau }  ).
\label{t-equi}
\end{equation}
Here, \( \tau \) represents the equilibration time, which directly reflects the rate of isospin diffusion. 
In Fig. \ref{8}(c), a linear relationship between the alignment angle \( \alpha \) and rotation time \( t \) is obtained. 
By substituting \( \alpha = kt \) into Eq. \ref{t-equi}, the neutron-proton equilibration can also be expressed as:
\begin{equation}
 (N-Z)/A = a+b\exp (-\frac{\alpha}{k\tau }  ).
\label{NZ-equi}
\end{equation}

These formulations show that when \( t \) is replaced with \( \alpha \), the rate constant encompasses not only the equilibration time \( \tau \) but also the linear relationship \( k \) between \( \alpha \) and \( t \). 
The parameter \( k \) exhibits significant dynamic effects and is positively related to \( K_0 \).

\section{\label{summary}CONCLUSION}

We focus on examining the effects of the incompressibility parameter \( K_0 \) on neutron-proton equilibration and constraining the density dependence of symmetry energy, taking into account the uncertainty in \( K_0 \). 
To achieve this, we employed the IQMD model, both with and without the GEMINI decay model, to simulate \( ^{70} \)Zn + \( ^{70} \)Zn collisions at 35 MeV/nucleon. 
The simulations demonstrate similar patterns to experimental data and reveal the dynamic mechanisms underlying the binary breakup of projectile-like fragments (PLFs).

Tracking the evolution of isospin asymmetry throughout the entire process of PLF binary breakup, we found that the isospin asymmetry of the HF $+$ LF system remains nearly constant over time, closely reflecting the isospin asymmetry of the original projectile. 
This indicates that the isospin asymmetry between the heavy fragment (HF) and light fragment (LF) is primarily governed by isospin diffusion, while the contribution from isospin drift, such as the emission of free neutrons or light neutron-rich fragments, is negligible.

The rotation time of PLFs is determined by identifying the moment of formation and subsequent breakup. It is observed that the rotation of PLFs is influenced by the transformation of angular momentum, which depends on the incompressibility \( K_0 \) applied in the model. 
In cases where \( K_0 \) is smaller, less relative angular momentum between the projectile and target is transferred into the angular momentum of the PLF, resulting in a smaller ratio between the alignment angle and rotation time. 
Conversely, for larger \( K_0 \), more angular momentum is transferred, producing the opposite effect. This ratio is directly tied to the equilibration rate constant \( c \), which explains why shifts in \( K_0 \) values affect the description of neutron-proton equilibration when it is measured by PLF rotation. 
Consequently, when using neutron-proton equilibration to constrain the density dependence of symmetry energy, the uncertainty in \( K_0 \) must be taken into account.

The calculations reveal that both a model with a smaller \( K_0 \) and a softer symmetry energy, and one with a larger \( K_0 \) and a slightly stiffer symmetry energy, can reproduce the experimental rate constant well. Taking into consideration the uncertainty in \( K_0 \), the slope of the symmetry energy at the saturation point is constrained within the range of \( L = 20 \sim 40 \) MeV. 
If the counterpart for isospin diffusion time lacks dynamic effects, the rate constant of neutron-proton equilibration would become independent of \( K_0 \), making it more sensitive to the density dependence of the symmetry energy.

\section*{ACKNOWLEDGMENTS}
This work was supported by the National Natural Science Foundation of China under Grants Nos. 12475136 and 12075327.

\bibliography{Esym_ref}
\end{document}